\documentclass[
PRL,
 preprint,
 showpacs,
 floatfix
]{revtex4-1}

\usepackage{epsfig}   

\newcommand{\C}{$\,^\circ$C}
\newcommand{\etal}[1]{{#1}~\emph{et al.}}
\newcommand{\micron}{$\mu$m}

\newcommand{\press}[2]{\mbox{${#1}\times10^{#2}$}}
\newcommand{\adir}{$[\bar{1}2\bar{1}]$}
\newcommand{\fig}[1]{Fig.~\ref{#1}}

\begin{document}

\title{The Equilibrium Shape of Graphene Domains on Ni(111)}
\author{Meifang Li$^1$, James B. Hannon$^2$, Rudoff M. Tromp$^2$,
        Jiebing Sun$^3$, Junwen Li$^4$, Vivek B. Shenoy$^4$, and Eric Chason$^1$}

\affiliation{$^1$School of Engineering, Brown University,
             Providence, Rhode Island 02912 \\
             $^2$IBM Research Division, T.J. Watson Research Center,
             Yorktown Heights, NY 10598 \\
             $^3$Department of Physics and Astronomy,
             Michigan State University, East Lansing, Michigan 48824, USA \\
             $^4$Department of Materials Science and Engineering,
             University of Pennsylvania, Philadelphia,
             Pennsylvania 19104, USA 
             }

%


\begin{abstract}
We have determined the equilibrium shape of graphene domains grown on Ni(111)
via carbon segregation at 925\C.
\emph{In situ}, spatially-resolved electron diffraction measurements
were used to 
determine the crystallographic orientation of the edges of the graphene
domains. In contrast to recent theoretical predictions of a 
nearly-circular shape, 
we show that graphene domains, which nucleate with random
shapes, all evolve toward a triangular equilibrium
shape with `zig zag' edges.  Only one of the two possible
zig-zag edge orientations is observed.
\end{abstract}




\maketitle

In many technological applications,  large,
defect-free graphene domains are
required in order to achieve high performance.  
Understanding the thermodynamics
and kinetics of graphene growth -- especially at metal surfaces -- is
consequently of great interest.  However, measuring the
properties of graphene domains during
growth is difficult due to the elevated substrate
temperature and harsh growth conditions.
Here we describe \emph{in situ} measurements
of the
\emph{equilibrium} shape of graphene domains grown on Ni(111) at
925\C. Our approach makes use of a novel imaging
mode that enables electron diffraction information to be obtained
from specific surface features 
during real-space imaging.

There have been several studies of the influence of growth
conditions on the shape of graphene 
domains.  For example, on Ni(111), \etal{Olle} 
showed that either hexagonal or triangular graphene
domain shapes
can be controllably synthesized under suitable
conditions~\cite{Oll12}.
On Cu(111), \etal{Chen} observed triangular
domains (with zig-zag edges) formed using ferrocene-dicarboxylic
acid as a carbon source~\cite{che12},
whereas conventional,
high-temperature CVD growth
using methane yields hexagonal domains with
zig-zag edges~\cite{Yu11}.  Clearly, growth kinetics can
play an important role in determining the graphene domain
shape.

Here we focus on the \emph{equilibrium} shape of graphene 
domains on Ni.
Ni(111) is a
particularly interesting substrate 
because graphene growth is \emph{epitaxial}.
As we show, the epitaxial relationship strongly influences the
thermodynamics of graphene domains.
Specifically, we find that graphene domains, which 
nucleate with random shapes, evolve toward triangular shapes 
that exhibit
only one type of zig-zag edge.  This is in contrast to
the hexagonal shape
found for free-standing
graphene~\cite{Gir09,Gan10}
as well as for graphene grown on (111) oriented 
Cu foils\cite{Yu11}. 
The observed triangular 
equilibrium shape is also in contradiction with a 
recent first-principles study~\cite{Art12}.  In that work the 
energy difference between the zig-zag and arm-chair orientations
on Ni(111) was found to be small, and a smooth, rounded
equilibrium shape was predicted~\cite{note1}.

In our experiments, graphene was grown on both crystalline
Ni(111) foils and
epitaxial Ni(111) films.  The films consisted of
100~nm of Ni sputter deposited onto annealed C-plane sapphire.
The foils were grown via
electrodeposition~\cite{Shi07}, with thicknesses in the
range 10 - 20~\micron.  Clean Ni(111)
surfaces were prepared by
\emph{in situ} sputtering with 1~keV Ne ions, followed
by annealing at 900\C.  Surface contamination was monitored using 
x-ray photoelectron spectroscopy.

\begin{figure}[h]
\epsfig{file=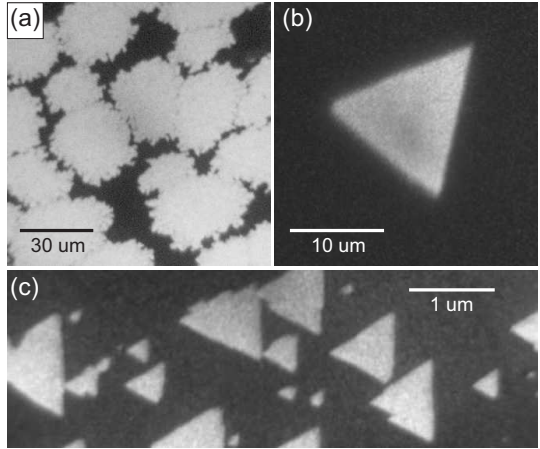, width= 2.8in} 
\caption{(a) PEEM image (850\C) recorded after the nucleation
of graphene domains in an ethylene pressure of
\press{5}{-7}~Torr. (b) PEEM image of a large, isolated triangular
graphene domain formed from the segregation of dissolved carbon. (c)
29~eV bright-field LEEM image of small graphene domains.}
\label{images}
\end{figure}

Graphene can be grown on Ni in two ways: direct 
CVD growth from the vapor, and by segregation of carbon
from the Ni bulk to the surface.
In our experiments, Graphene was first
grown by exposing the surface to \press{5}{-7}~Torr of
ethylene at
a substrate temperature of 850\C.  During ethylene exposure, the surface
was imaged using photoelectron emission spectroscopy (PEEM).
When graphene nucleation was observed, the 
ethylene pressure was reduced.  Because of
the relatively high growth rate, the
domain shape
was irregular and clearly not equilibrated, as shown 
in \fig{images}a.  Better control
of the graphene growth rate was achieved using cycles of dissolution and 
segregation.  Upon heating to ~990\C, graphene domains dissolve
as carbon migrates to the bulk due to increased carbon solubility
at elevated temperature.  When the temperature is lowered (e.g.\ to
925 \C) graphene domains reform as carbon segregates to the surface
as shown in \fig{images}b,c.

The growth rate during segregation can be controlled by adjusting the
substrate temperature, and can be directly monitored
using LEEM.  If the growth rate is low, large 
triangular domains are observed, despite the fact that a
typical domain crosses many tens of steps
(the average terrace width is roughly $100$~nm).  This shows
that atomic-height surface steps do not strongly influence the 
shape of the domains.
However, when the surface morphology is 
significantly disrupted, e.g.\ at twin boundaries or
at  large step bunches,
the domain boundary can be influenced by the local surface morphology.

To equilibrate the domain shapes, we used 
\emph{in situ} LEEM imaging, and the ability to precisely control
the growth rate, to `anneal' the domains at essentially constant area.
We grew isolated domains
at a low growth rate ($\approx$0.2~\micron$^2$/s) until the areas reached
roughly 1~\micron$^2$.  The temperature was then increased slightly
so that the
domains ceased to grow.  The shapes were then observed for 
an extended period of time.  We found that the domains all adopted 
polygonal shapes with edges
normal to \adir\ (\fig{images}c).  Over
time, the domains evolved towards a simple
triangular shape (\fig{images}b), indicating that this shape
is the equilibrium geometry.
%
%
%
%
  
The graphene domains exhibit a $1\times1$ diffraction
pattern in 
spatially-resolved LEED.  The
diffraction pattern indicates that the graphene is epitaxial,
as expected, but
does not directly reveal the orientation of the domain edge. 
To determine the crystallographic orientation of the 
substrate, the diffraction pattern must be oriented
with respect
to the real-space image. To do this we exploit the 
spherical aberration of the cathode lens system to generate
a `real-space' diffraction pattern.

Due to diffraction at the sample surface, the incident
electron beam in LEEM generates diffracted beams, in addition
to the specularly-reflected beam. 
These beams  emerge
from the sample at specific angles defined
by the diffraction condition.  In a perfect 
imaging system (free of spherical aberration) the
images formed from the different beams will coincide 
in the image plane.  However in LEEM the spherical aberration
of the cathode lens is  large,
and the images arising from different diffracted beams
do not perfectly coincide, but are displaced with respect
to each other in manner that reflects the crystal
structure and symmetry. In fact, measuring the 
relative positions of these images can be used to 
determine the spherical aberration coefficient~\cite{Tro13}.

Here we use the real-space diffraction pattern formed by 
the displaced images to determine the orientation of the 
graphene domains with respect to the Ni reciprocal lattice,
and hence the crystallographic orientation of the domain edge.
%
%
\begin{figure}[h]
\epsfig{file=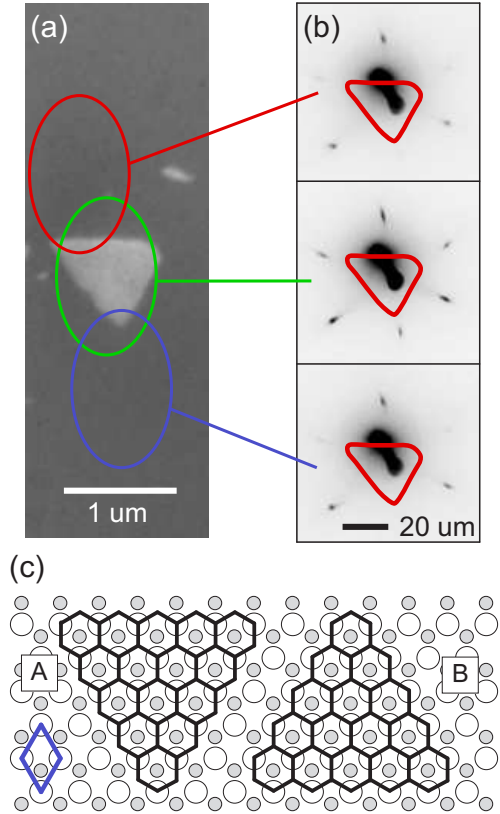, width= 2.6in} 
\caption{(a) 12~eV bright-field
LEEM image of an isolated graphene domain. (b)
47~eV real-space
LEED images,
40$\times$40~\micron$^2$,
recorded from the oval areas indicated in (a).  The
domain edges are perpendicular to Ni reciprocal
lattice vectors, indicating that the edge have
the `zig-zag' orientation. (c) Graphene domains
with $A$ and $B$-type unreconstructed zig-zag edges.  The
Ni(111) unit cell is indicated in blue. }
\label{rsleed}
\end{figure}
%
%
An image of an isolated domain is shown in 
\fig{rsleed}a.  Real-space diffraction images were recorded
using a small aperture ($\sim$1~\micron\ diameter) to illuminate 
specific regions of the surface near the domain.  The 
 areas illuminated by the incident beam
 are indicated by ovals in \fig{rsleed}a, 
and the corresponding real-space diffraction patterns are
shown in \fig{rsleed}b.  In the \fig{rsleed}b, each diffracted
beam produces a $\sim$1\micron\ image of the illuminated
area. When the Ni is illuminated, the pattern is clearly
three-fold symmetric.  When the graphene domain is illuminated, 
the pattern is still three-fold symmetric, but all six first-order
diffraction spots are visible.  Comparison with the domain
shape (shown as a red outline) shows that the domain edges
are perpendicular to the reciprocal lattice vectors of the 
substrate.  This, in turn, shows that the graphene edges
are parallel to the
`zig-zag' orientation (normal to \adir).

For free-standing graphene, or graphene on incommensurate 
substrates including Cu(111), the six possible zig-zag orientations
are equivalent, and the equilibrium shape is hexagonal.  
However, because graphene on Ni(111) is epitaxial, there
are two inequivalent zig-zag orientations as shown
in \fig{rsleed}c, which are labeled
$A$ and $B$.  The local environments of the carbon edge atoms are
very different in the $A$ and $B$ structures, at least for the 
un-relaxed, un-reconstructed structures shown in \fig{rsleed}c.
The edge atoms in $A$ are located over \emph{fcc} hollow sites,
whereas the edge atoms in the $B$ are located directly on top of
substrate Ni atoms.

To determine which type of edge is observed in experiment
(i.e. $A$ or $B$),
the intensities of the diffracted spots must be considered.
The six first-order diffraction spots are not equivalent. 
As shown in \fig{rsleed}b (top panel), for diffraction
from Ni(111) at 47~eV, three spots
are bright and three are very dim.  
The intensities
of the diffracted beams are determined by the 
positions of the Ni atoms within the unit cell, and can be computed
for various crystal structures~\cite{Pen74}.
%
%
\begin{figure}[h]
\epsfig{file=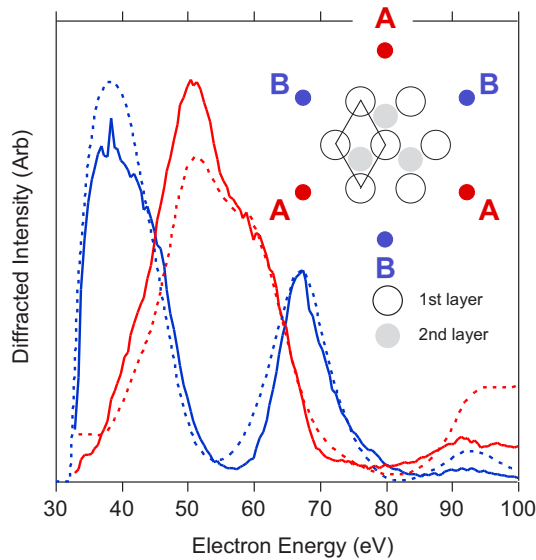, width= 2.8in} 
\caption{Measured (solid) and compute (dashed) diffracted
intensity versus electron energy for the two inequivalent first-order
diffraction beams, labeled $A$ and $B$, of Ni(111).  The inset shows the
orientation of the diffracted beams with respect to the (111) surface. 
The Ni(111) unit cell is indicated.}
\label{iv}
\end{figure}
%
%
The dashed lines in \fig{iv}
show the calculated diffracted intensity for the $A$ and $B$ beams as
a function of the
incident electron energy~\cite{Sun10}.  The calculation shows
that for an energy of 40~eV, the $B$ diffraction spots
will be significantly more intense than the $A$ spots.
At 50~eV, the situation is reversed, and the $A$ spots will
be intense.   Intensity-versus-voltage (or `IV') curves
measured in this work are shown as solid lines in \fig{iv}
and agree well with calculated intensities (dashed lines).

The orientation of the diffracted beams with respect to the
Ni(111) unit cell is indicated in the inset to \fig{iv}.
This information can be used to determine the orientation of the 
crystal in \fig{rsleed}.  The diffraction patterns from
Ni(111) shown in
\fig{rsleed}b (top and bottom panels) were recorded
at 47~eV, and therefore the bright
spots are the $A$ diffracted beams of \fig{iv}.  The
corresponding crystal orientation is shown in \fig{rsleed}c.
Once the crystal orientation is determined, it is clear that
the observed domain shape has the $A$-type zig-zag structure.

The favorability of the $A$-type edge is, perhaps, puzzling.
In the unrelaxed structure, the carbon edge atoms 
are located over  \emph{fcc} hollow
sites.  In contrast, the edge atoms in the $B$-structure 
are located directly over Ni atoms, and presumably interact
strongly with the Ni substrate.
Furthermore, the  triangular shape contradicts
the expectation of a compact, rounded domain shape based on
recent first-principles calculations~\cite{Art12}.
In that work, \etal{Artyukhov} considered a number of 
edge terminations for graphene domains on several metal
surfaces.  For Ni(111), the formation energies of the 
zig-zag edge and reconstructed 
arm-chair edge (A5') were found to be roughly
0.4~eV/\AA.  A reconstructed version of the zig-zag
edge (Z57) was found to have considerably higher 
energy than the unreconstructed zig-zag structures
shown in \fig{rsleed}c.
The similarity of the edge energies suggests
a nearly circular equilibrium shape, especially at 
elevated temperature.  One explanation for the 
clear stability of the 
$A$-type zig-zag orientation that we observe is that the 
edge structure is more complicated than the
structures
considered by \etal{Artyukhov}.  Specifically, those
authors only considered reconstructions that involved
a rearrangement of the carbon atoms at the edge. 
Given the significant interaction between graphene and 
Ni, as well as the high mobility of Ni atoms during
graphene growth, it may be that the equilibrium edge
structure incorporates Ni.  To test this hypothesis
we computed the formation energies of 
several edge structures that incorporate excess Ni.

The calculations were carried out by using the
Vienna \emph{ab initio} simulation package (VASP),\cite{Kre96} 
with exchange-correlation functional described by Ceperley-Alder local
density approximation (LDA)\cite{Cep80} and interaction
between core electrons and valence electrons by the frozen-core
projector-augmented wave (PAW) method.\cite{Blo94} An
energy cutoff of 415~eV was used for the plane wave basis expansion.
The Ni(111) substrate was modeled by slabs of four atomic layers.
A vacuum region of more than 12~\AA\ was used to prevent the
interaction between periodic slabs.  During the structural relaxation,
the Ni atoms of bottom two layers were fixed in their bulk position.
All other atoms were allowed to relax until the force on each
atom was smaller than 0.02~eV/\AA.\  A 1$\times$12$\times$1 $k$-point
scheme was used to sample reciprocal space.

Graphene ribbons with a width of 
$\sim16$~\AA\ were used to compute the edge energies.
The ribbon supports
$A$ and $B$-type edges on opposite sides.
We find that the
zig-zag edge formation energy, e.g.\ $(E_A+E_B)/2$, is
0.36~eV/\AA, in good agreement with
\etal{Artyukhov}~\cite{Art12}.
We considered a number of $A$-type edge terminations that 
involve more complex Ni-C edge structures, including: (1) 
Ni adatoms at the graphene edge, and (2) Ni atoms substituted for
C edge atoms.  None of the structures we considered has a lower
formation energy than the unreconstructed zig-zag structures shown in
\fig{rsleed}c.  Structures
with Ni substituted for C are particularly unfavorable.  It 
is possible that the actual structure is more complicated than the ones
we considered. In addition, the elevated growth temperature
could be an important factor.  The calculations
correspond to zero temperature,
and it may be that entropy plays a significant role in
stabilizing a Ni-decorated structure at the growth temperature
(925\C).  That is, the equilibrium shape at high temperature could be
very different from that at low temperature.

In summary, we have used a novel real space diffraction measurement
technique to determine the orientation of equilibrated graphene 
domains on Ni(111). At 925\C, the random domain
shapes observed
after nucleation
evolve towards a triangular shape with the 
$A$-type zig-zag edges.   The fact that
the $A$-type orientation is favored is in contraction with
first-principles calculations that suggest that the 
energy difference between zig-zag and arm-chair orientations
is small.
It may be that triangular shape we observe is stabilized
by a more complex C-Ni edge structure, which could
explain the apparent discrepancy with
earlier theoretical work.


\begin{thebibliography}{14}
\bibitem [{not()}]{note1}%
  \BibitemOpen
  \href@noop {} {}\bibinfo {note} {The authors find that the lowest energy
  armchair orientation is actually reconstructed.}\BibitemShut {Stop}%
\end{thebibliography}
%

\end{document}